\documentclass[review]{elsarticle}

\usepackage{amsfonts}
\usepackage{epsfig}
\usepackage[font=small,labelfont=bf]{caption}
\usepackage{amssymb,amsmath,graphicx,latexsym}

\usepackage{graphicx}
\usepackage{caption}
\usepackage{subcaption}

\usepackage{bm}
\usepackage[table]{xcolor}

\usepackage{caption}
\newtheorem{theorem}{Theorem}
\newtheorem{remark}{Remark}

\usepackage[
top    = 3.50cm,
bottom = 3.50cm,
left   = 3.2cm,
right  = 3.2cm
]{geometry}

\usepackage{lipsum}
\makeatletter
\def\ps@pprintTitle{%
 \let\@oddhead\@empty
 \let\@evenhead\@empty
 \def\@oddfoot{}%
 \let\@evenfoot\@oddfoot}
\makeatother

\usepackage{xcolor}

\usepackage{soul}

\begin{document}

\begin{frontmatter}

\title{ MLE for  the parameters of  bivariate  interval-valued  model \\ 
}    

\date{ }

\author[address1]{S. Yaser Samadi\corref{mycorrespondingauthor}}
\cortext[mycorrespondingauthor]{Corresponding author}
\ead{ysamadi@siu.edu}

\author[address2]{L. Billard}

\author[address3]{Jiin-Huarng Guo}

\author[address4]{Wei Xu}

\address[address1]{Department of Mathematics, Southern Illinois University, Carbondale IL 62901, USA}
\address[address2]{Department of Statistics, University of Georgia,  Athens, GA 30602, USA}
\address[address3]{Department of Applied Mathematics,  National Pingtung University,  Taiwan, R. O. C.}
\address[address4]{Capital One, McLean VA 22102 USA.}

\begin{abstract}

  With contemporary data sets becoming too large to analyze the data directly, various forms of aggregated data are becoming common. The original individual data are points,   but after aggregation the observations  are interval-valued (e.g.).  While some researchers simply analyze the set of averages of the observations by aggregated class, it is easily established that  approach ignores much of the information in the original data set. The initial theoretical work for interval-valued data was that of Le-Rademacher and Billard (2011),  but those  results were limited to estimation of the mean and variance of a single variable only. This article seeks to redress the limitation of their work by deriving the maximum likelihood estimator for the all important covariance statistic, a basic requirement for numerous methodologies, such as  regression, principal components, and canonical analyses.  Asymptotic properties of the proposed estimators are established.  The Le-Rademacher and Billard results emerge as special cases of our wider derivations.
 \end{abstract}


\end{frontmatter}

\section{Introduction}
 Le-Rademacher and Billard (2011) derived the maximum likelihood estimator (MLE) of the mean and variance when the data consisted of interval-valued observations.
In this work, our primary focus is to extend that work   to derive the MLE for the covariance function between the  random variables $(X, ~Y)$ when both $X$ and $Y$ are interval-valued. The covariance function is of particular importance given its  major role in  covariance matrices used in principal component analyses, canonical analyses, as well as the entries/components in the ``${\bf X}'{\bf X}$" and ``${\bf Y}'{\bf X}$" terms in multiple regression analyses, among other methodologies; yet so far no maximum likelihood estimator currently exists for this key entity.

Maximum likelihood estimation is arguably the most important method of parameter estimation in statistics.
 Traditionally, the data  analyzed are so-called classical data in which each data value  is a single point. However, with the advent of modern computers and data collecting devices, contemporary data sets are becoming too large and too complicated to be analyzed directly. One approach has been to aggregate the observations into classes or categories of observations,    
 with the  nature of the aggregation  varying depending on the underlying scientific questions.  
  One consequence of this phenomena is that the data  to be analyzed are  no longer    single points. Instead, they are so-called symbolic data first identified by Diday (1988). Thus, rather than the points of classical data, symbolic data are hypercubes (or Cartesian products of distributions). That is, for each variable, realizations  can be intervals, lists, histograms, or so on. For example, an interval realization can be $[10, 30]$ in contrast to a classical point observation $20$. One distinguishing feature of such data is that realizations (e.g.,  intervals) have internal variations that are not present in the point values of classical data. Taking classical surrogates (e.g., interval midpoints) ignores these internal variations and in doing so throws away valuable information inherent to the observations thus producing results that are not  necessarily  correct.
  Many methodologies transform the single interval-valued $X=[a,b]$ say, into two  variables $X^c=(a+b)/2$ and $X^r=(b-a)/2$, center and range variables (or equivalently into $X_a=a$ and $X_b=b$, the two end points).  While this is an  improvement on using $X^c$ only, the resulting analyses can give inaccurate answers; as illustrated in  Appendix A. Recently, Oliveira et al. (2017, 2022) proposed a model, based on  functions of the  centers and ranges, linking the micro-data (un-aggregated observations) that made up the macro-data (interval) and thence looked at eight possible formulations in an effort to tie together proposed covariance functions from the literature.

 Many methodologies for symbolic datasets have been developed since Diday (1988) introduced the concept. By and large, these are intuitive extensions of their classical counterparts, although accommodating the internal variations of symbolic data make these extensions far from trivial. However, since classical data are a special case of symbolic data (e.g., the classical value $x$ is equivalent to the  interval value $[x,\, x]$), obtaining the established classical results as a special case of the symbolic techniques is one form of verification as to their correctness. Deriving theoretical foundations for these methodologies has been more difficult to achieve. Some results do however exist. Thus, Diday (1995), Diday, Emilion and Hillali (1996), Emilion (1997), and Diday and Emilion (1996, 1998) obtained a mathematical framework for some classes of symbolic data. Then, Diday and Emilion (1998, 2003) and Brito and Polaillon (2005) extended those results to Galois lattices used in, e.g.,  pyramid clustering. Later, Le-Rademacher and Billard (2011) obtained maximum likelihood estimators for the mean and variance of a univariate distribution when the observations were interval-valued. More recently, Zhang et al. (2020) and Beranger et al. (2022) provided a framework for likelihood functions for interval data, which results were then used to consider the variance-covariance function in regression; see Whitaker et al. (2020, 2021), and Rahman et al. (2020). Samadi and Billard (2021) looked at auto-covariances for interval-valued time series observations. Beyond these results, almost no mathematical theoretical underpinnings for symbolic methodology have  yet been fully established.

First in Section \ref{Sec:2},  we describe symbolic data especially interval-valued data, along with some of the basic descriptive statistics for these data.  Then in Section \ref{Sec:3}, we establish a likelihood function, and derive in Section 4.1 some maximum likelihood estimators for the parameters of the proposed model.  Hence, by utilizing the ideas behind first and second order conditional moments, we  show that the empirically based descriptive statistics (of Section \ref{Sec:2}) are maximum likelihood estimators (or, approximately so); see Section \ref{Sec:4.2}. Asymptotic properties of proposed estimators  are studied  in Section \ref{Sec:asympto}.
 In Section \ref{Sec:6}, we  consider some  extensions where the internal  spread  across the intervals is non-uniform.
Then, in Section \ref{Sec:7}, a short simulation study is conducted; and the results are illustrated through an analysis of real data.

 \section{Symbolic Data}\label{Sec:2}
A major source of symbolic data arises when managing large contemporary data sets. For example, a medical insurer may have a database of client visits to a health-care entity (doctor, hospital, etc.) with entries recording values for a variety of variables, such as basic medical information (e.g., weight, cholesterol, blood pressure, \dots), or demographic information (gender, age, \dots), disease diagnosis (e.g., cancer with a list of cancers presenting, heart disease with a list of heart conditions, and so on), clinical diagnostics, geographical variables, and more. The insurer (or investigator, or \dots) is not so much interested in a particular visit to the doctor as s/he might be interested in the health characteristics of categories of clients.
However, symbolic data can also occur naturally. For example, the pileus cap width of mushroom species is usually recorded as intervals;  interval data can also be used to protect confidentialities; daily temperatures are recorded as minimum-maximum temperatures, e.g., $[38, 56]$.   Bock and Diday (2000) and Billard and Diday (2003, 2006) have extensive detailed descriptions and examples of symbolic-valued data, including  intervals.

It is important to remind ourselves that although observations may be aggregated  into intervals (say), the underlying distributions are still the traditional (classical) distributions with their relevant parameters.
That is,  after aggregation, the   original classical values still retain their same distributions. What has changed is the format of the realization of the observation to become, e.g., an interval-valued observation.   Indeed, Bertrand and Goupil (2000) obtained the sample mean and the sample variance for a sample of interval-valued observations as  point estimators of the parameters $\mu$ and $\sigma^2$.    For the symbolic-valued interval,  additional assumptions are made as to how those aggregated observations are spread within a given interval however. The Bertrand and Goupil results assume the observations are uniformly spread across the intervals.  
 As an aside, we note this uniformly assumption is comparable to that used in finding the histogram of grouped  data  as taught in elementary statistics  classes.  We note that this is not the same as interval-arithmetic concepts (which would give parameters themselves as interval-values, e.g., $\mu = (\mu_1, \mu_2)$; see, e.g., Moore, 1966).  The interval-arithmetic domain is quite different from the symbolic-valued data domain, and is not considered herein.

Interval-valued  data are, as the name suggests, realizations of a random variable $Y$,  that are intervals $[a_i,~b_i]$, $a_i \leq b_i$, $i = 1, \dots, n$, for a random sample of size $n$. [Intervals can be open or closed at either end.] Now, the observable interval is essentially an (aggregated) group of observations, while the unobservable individual data are scalar.   Therefore, the underlying parameters and descriptive statistics, such as the sample mean and variance,  are also scalar entities.  Bertrand and Goupil (2000) first obtained expressions for the empirical (sample) mean $\bar Y$ and variance $S^2_Y$ as  
\begin{align} \label{eq1}
 \bar Y &= \frac{1}{2n}\sum_{i=1}^n(a_i + b_i),\\ \label{eq2}
S^2_Y &= \frac{1}{3n}\sum_{i=1}^n(a_i^2+a_ib_i+b_i^2) - \bar{ Y}^2/n \equiv S_{YY} \mbox {, say}.
\end{align}

Let us also have a second random variable $X$ with interval-valued realizations $[c_i,\, d_i]$, $i = 1,\dots, n$, for a random sample of size $n$.  Billard (2008) obtained the empirical sample covariance function as
\begin{align}\label{eq3}
\begin{split}
S_{XY}&=\frac{1}{6n}\sum_{i=1}^{n}[2(a_{i}-\bar{Y})(c_{i}-\bar{X})+(a_{i}-\bar{Y})(d_{i}-\bar{X})  \\
&~~~~~~~~~~~~~~~~~~~~~~~~+(b_{i}-\bar{Y})(c_{i}-\bar{X})+2(b_{i}-\bar{Y})(d_{i}-\bar{X})]
\end{split}
\end{align}
where $\bar Y$ is as given in Eq.($\ref{eq1}$) and $\bar X = \frac{1}{2n}\sum_{i=1}^n(c_i + d_i)$.
When $Y = X$, then Eq.($\ref{eq3}$) becomes $S_{XX} \equiv S_{X}^2$. For the special case of classical data where $a_i = [a_i,\, a_i]$ and $c_i = [c_i,\, c_i]$, the formulas in Eqs.($\ref{eq1}$)-($\ref{eq3}$) all reduce to their well-known classical counterparts.

An implicit assumption in the derivation of these empirical results is that observations within intervals are uniformly spread across the interval. Billard (2008) showed how the results can be extended to the non-uniform case illustrating with a triangular distribution. Since however most methodology for interval data are based on this uniformly spread assumption, our maximum likelihood approach that follows will retain this assumption.
 Cariou and Billard (2015) provides a test for this basic but important assumption.

\section{Symbolic Likelihood Function}\label{Sec:3}
As in the preceding sections, suppose we have two random variables ($X, Y$) with interval-valued realizations ($X_i, Y_i$) where
$X_i = x_{i} = [c_i, d_i]$ and $Y_i = y_{i} = [a_i, b_i]$ for $i = 1, \dots, n$.
Let ($X_i,\, Y_i$) have joint (bivariate) probability density function (pdf) $h_{X,Y}(x,y; \mbox{\boldmath$\delta$})$ with parameter $\mbox{\boldmath$\delta$}$. Classes of inferential results when  realizations of $(X,\, Y)$ are classical point values in $\mathbb{R}^2$ are well established (see, e.g., Casella and Berger, 2002; Lehmann, 1983, 1986). Our focus however is for interval-valued realizations. We adapt the approach of Le-Rademacher and Billard (2011) for univariate interval-valued random variables.

Since each variable has aggregated  observed values over an interval, we need to consider the internal distribution of those values within the interval. Therefore, let the joint and marginal internal distributions be defined by, respectively, for $i = 1,\dots,n$,
\begin{equation}\label{eq5}
\mbox {Within  } (X_i, Y_i)  \sim f_i^{xy}(x_{i}, y_i; {\bf \Theta}_i^{xy}),\mbox {Within  } (X_i)  \sim f_i^{x}(x_{i}; {\bf \Theta}_i^{x}),\mbox {Within  } ( Y_i)  \sim f_i^{y}(y_i; {\bf \Theta}_i^{y}).
\end{equation}

These distributions $f_i^{xy}$, $f_i^x$ and $f_i^y$  along with their parameters ${\bf \Theta}_i^{xy}$,  ${\bf \Theta}_i^x$ and ${\bf \Theta}_i^y$ are internal (or ``within" observation) entities, distinct from the overall distributions $h_{X,Y}(x,y; \mbox{\boldmath$\delta$})$. Consistent with current symbolic data analyses methodology for interval observations, suppose the internal distributions within the ($X, Y$) intervals, i.e., the $f_i^x(x_{i}; {\bf \Theta}_i^x)$ and $f_i^y(y_{i}; {\bf \Theta}_i^y)$ of Eq.($\ref{eq5}$), are uniformly distributed, for each $i = 1,\dots,n$. Hence, for the intervals $Y_i $, realizations of $\Theta_{i1}^y$ are $\theta_{i1}^y = (a_i+b_i)/2$; and likewise for the intervals $X_i $ realizations of $\Theta_{i1}^x$ are $\theta_{i1}^x = (c_i+d_i)/2$. The variation variables $\Theta_{i2}^x, \Theta_{i2}^y, \Theta_{i2}^{xy}$, respectively,  have realizations $\theta_{i2}^y = (b_i-a_i)^2/12$, $\theta_{i2}^x = (d_i-c_i)^2/12$ and $\theta_{i2}^{xy} = (b_i-a_i)(d_i-c_i)/12$.  

Also, since $(X_i, Y_i), ~i = 1,\dots,n,$ are random variables, the parameters $({\bf \Theta}_i^x, {\bf \Theta}_i^y)$ are not fixed, taking different values as $(X_i, Y_i)$ change with $i = 1,\dots,n$. That is, $({\bf \Theta}_i^x, {\bf \Theta}_i^y)$ are themselves random variables. Therefore, let the underlying distribution of $({\bf \Theta}_i^x, {\bf \Theta}_i^y)$ be
\begin{equation}\label{eq6}
({\bf \Theta}_i^x, {\bf \Theta}_i^y)\sim g_{xy}({\bf \Theta}_i^x,{\bf \Theta}_i^y ; \mbox{\boldmath$\tau$}_{xy}), ~ i = 1,\dots,n.
\end{equation}

Then for these parametric families $g_{xy}({\bf \Theta}_i^x, {\bf \Theta}_i^y; \mbox{\boldmath$\tau$}_{xy})$ with parameter $\mbox{\boldmath$\tau$}_{xy}$, there exist one-to-one correspondences between the $(X_i, Y_i)$ and $({\bf \Theta}_i^x, {\bf \Theta}_i^y)$. To illustrate this one-to-one correspondence, for the uniform case, assume that $X$ (similarly, for $Y$) has realization
\begin{itemize}
\item[(i)] $x = [c, d] = [1,3]$; then $\theta_i^x =2$, $\theta_2^x = 1/3$.
\item[(ii)] If $\theta_i^x = u$, $\theta_2^x = \omega > 0$, then $c + d= 2u$ and $(d-c)^2 = 12 \omega$.
\end{itemize}
It can be shown that for $c < d$, we have $c = u -  \sqrt{3\omega}$, $d=u +  \sqrt{3\omega} $. When $u = 2$, $\omega = 1/3$, we obtain $x = [c,d] = [1, 3]$, as required.

Therefore, because of this one-to-one correspondence,
 we have
\begin{equation} \label{eq7}
h_{X,Y}(x_i, y_i; \mbox{\boldmath$\delta$}) = g_{xy}({\bf \Theta}_i^x, {\bf \Theta}_i^y; \mbox{\boldmath$\tau$}_{xy}).
\end{equation}

We note that the parameters $\mbox{\boldmath$\delta$}$ relate to the overall distribution $h_{X,Y}(x,y; \mbox{\boldmath$\delta$})$, and the parameters $\mbox{\boldmath$\tau$}_{xy}$ relate to  the internal distributions of the (${\bf \Theta}_i^x, {\bf \Theta}_i^y$) variables. Therefore, given the one-to-one correspondence between $(X_i, Y_i)$ and $({\bf \Theta}_i^x, {\bf \Theta}_i^y)$,  then there is a one-to-one correspondence between  $\boldsymbol\delta$ and $\boldsymbol\tau_{xy}$. The components of this $\boldsymbol\tau_{xy}$ will depend on the internal variation distributions $f_i^{x}(x_{i}; {\bf \Theta}_i^{x})$ and $f_i^{y}(y_i; {\bf \Theta}_i^{y})$ as shown in the next Section 4.1.

 We can write the likelihood function of the parameter $\mbox{\boldmath$\delta$}$ given the data $(X_i, Y_i) = (x_{i}, y_{i}),~ i = 1,\dots,n,$ as
\begin{equation} \label{eq8}
L(\mbox{\boldmath$\delta$}; (x_{1}, y_1),\dots,(x_{n},y_{n})) = \prod_{i=1}^nh_{X,Y}(x_i,y_i; \mbox{\boldmath$\delta$});
\end{equation}
and substituting from Eq.($\ref{eq7}$), we have
\begin{equation}\label{eq9}
L(\mbox{\boldmath$\delta$}; (x_{1}, y_1),\dots,(x_{n},y_{n})) = \prod_{i=1}^ng_{xy}(\mbox{\boldmath$\theta$}_i^x, \mbox{\boldmath$\theta$}_i^y; \mbox{\boldmath$\tau$}_{xy}) = L(\mbox{\boldmath$\tau$}_{xy}; \mbox{\boldmath$\theta$}_1^x,\dots,\mbox{\boldmath$\theta$}_n^x, \mbox{\boldmath$\theta$}_1^y,\dots, \mbox{\boldmath$\theta$}_n^y)
\end{equation}
where $\mbox{\boldmath$\theta$}_i^x, \mbox{\boldmath$\theta$}_i^y$ are realizations of $\mbox{\boldmath$\Theta$}_i^x, \mbox{\boldmath$\Theta$}_i^y$, $i = 1, \dots, n$.

Note that when there is independence, the product of the marginal distributions can replace the joint distribution $g_{xy}(\mbox{\boldmath$\Theta$}_i^x, \mbox{\boldmath$\Theta$}_i^y; \mbox{\boldmath$\tau$}_{xy})$ in Eq.($\ref{eq7}$), i.e., we have for $\mbox{\boldmath$\Theta$}_i^x$ and $\mbox{\boldmath$\Theta$}_i^y$, respectively,
\begin{equation}{\nonumber}
\mbox{\boldmath$\Theta$}_i^x \sim g^x(\mbox{\boldmath$\Theta$}_i^x; \mbox{\boldmath$\tau$}^x), ~~~ \mbox{\boldmath$\Theta$}_i^y \sim g^y(\mbox{\boldmath$\Theta$}_i^y; {\bf \tau}^y),~~i=1,\dots,n.
\end{equation}

 Since the data $( \mbox{\boldmath$\theta$}_1^x,\dots, \mbox{\boldmath$\theta$}_n^x,  \mbox{\boldmath$\theta$}_1^y,\dots, \mbox{\boldmath$\theta$}_n^y)$ in Eq.($\ref{eq9}$) are now classically valued observations, we can apply  maximum likelihood methods for classical data to  estimate the parameters of interest.  This is done in  Section 4.

\section{Maximum Likelihood Estimators}
In Section 4.1, we obtain the maximum likelihood estimators for the within observation parameters. Then in Section 4.2, these are used to obtain the overall (within plus between) variation estimators.

\subsection{Estimators for Within Observation Parameters}
Let us take the internal parameters ${\bf \Theta}_i^x$ associated with $X_i$ and ${\bf \Theta}_i^y$ associated with $Y_i, ~i = 1,\dots,n,$ introduced in Section 3  as   ${\bf \Theta}_{i1} = (\Theta_{i1}^x, \Theta_{i1}^y)$ and ${\bf \Theta}_{i2} = (\Theta_{i2}^x, \Theta_{i2}^y, \Theta_{i2}^{xy})$. That is, we let   the $\Theta_{i1}^x$ and $\Theta_{i1}^y$ correspond to the internal mean of $X_i$ and $Y_i$, respectively, for each $i = 1,\dots,n$; and the $\Theta_{i2}^x$ and $\Theta_{i2}^y$ correspond to the internal variation of $X_i$ and $Y_i$, respectively, and $\Theta_{i2}^{xy}$ corresponds to the covariance of $(X_i,Y_i)$ for each $i = 1,\dots,n$.
At this stage, it is necessary to specify distributions governing these internal parameters. It is not unreasonable to consider the corresponding conjugate distribution as the respective distribution. Therefore, if the underlying individual observations (i.e., non-aggregated data)   are assumed to be normally distributed (or asymptotically so), then  the conjugate distribution for the internal  mean parameters is normally distributed. Likewise, given its conjugate  role for covariance functions and its associated property of being non-negative definite and symmetric,  the Wishart (1928) distribution is used for the internal variance-covariance parameters. However, other underlying distributions could be considered.

  Formally, for each $i = 1,\dots,n,$     suppose that the joint distribution of the internal means ${\bf \Theta}_{i1} = (\Theta_{i1}^x, \Theta_{i1}^y)$ is a bivariate normal distribution $N_2(\mu_x, \mu_y, \sigma_x^2, \sigma_y^2, \rho)$;
and  suppose ${\bf \Theta}_{i2} = (\Theta_{i2}^x, \Theta_{i2}^y, \Theta_{i2}^{xy})$ follows a bivariate Wishart distribution 
(originally derived by Fisher, 1915) defined by
\begin{equation}\label{eq10}
  \begin{split}
    f(\theta_{2}^{x}, \theta_{2}^{y} , \theta_{2}^{xy}; \gamma_1,\gamma_2,\gamma_3 ) &=  \frac{  \left(\gamma_1\gamma_2-\gamma_3^2 \right)^{-\frac{\nu}{2}}  }{2 \sqrt[\nu]{\pi}\Gamma(\frac{\nu}{2}) \Gamma(\frac{\nu-1}{2})}   \left(\theta_{2}^{x}\theta_{2}^{y} -(\theta_{2}^{xy})^2 \right)^{\frac{\nu-3}{2}} \\
     &~~~~~~~~~~ \times  \exp\left\{-\frac{\gamma_1\gamma_2}{2(\gamma_1\gamma_2-\gamma_3^2)} \left( \frac{\theta_{2}^{x}}{\gamma_1}+\frac{\theta_{2}^{ y}}{\gamma_2}-\frac{2\gamma_3 \theta_{2}^{ xy}}{\gamma_1 \gamma_2}\right)\right\}
  \end{split}
\end{equation}
 where  $\nu>2$  is the degree of freedom,   $\theta_{2}^{x}>0$, $\theta_{2}^{ y}>0$, and $ - \sqrt{\theta_{2}^{x} \theta_{2}^{ y}}  <\theta_{2}^{ xy}<   \sqrt{\theta_{2}^{x}\theta_{2}^{y}}$ (see Anderson, 2003).
 Then, the joint probability density function $g_{xy}({\bf \Theta}_i^x,{\bf \Theta}_i^y ; \mbox{\boldmath$\tau$}_{xy})$ of Eq.($\ref{eq6}$) can be written as, with $\mbox{\boldmath$\tau$}_{xy} = (\mu_x, \mu_y, \sigma_x^2, \sigma_y^2, \rho, \gamma_1,\gamma_2,\gamma_3)$,
\begin{equation}\label{eq11}
g_{xy}({\bf \Theta}_{i1},{\bf \Theta}_{i2} ; \mbox{\boldmath$\tau$}_{xy}) = g_1(\Theta_{i1}^x, \Theta_{i1}^y; \mu_x, \mu_y, \sigma_x^2, \sigma_y^2, \rho) \times g_2(\Theta_{i2}^x, \Theta_{i2}^y, \Theta_{i2}^{xy}; \gamma_1,\gamma_2,\gamma_3).
\end{equation}

Let us write the likelihood function based on the intervals, $L_I$, from Eq.($\ref{eq9}$) with ${\bf \Theta}_n$ representing the observations, as
\begin{align*}
L_I &\equiv L_I(\mbox{\boldmath$\tau$}; {\bf \Theta}_n) \equiv L_I(\mu_x, \mu_y, \sigma_x^2, \sigma_y^2,\rho,\gamma_1,\gamma_2,\gamma_3;  \\
& ~~~~~~ {} \theta_{11}^x,\dots,\theta_{n1}^x, \theta_{11}^y,\dots,\theta_{n1}^y,  \theta_{12}^x,\dots,\theta_{n2}^x, \theta_{12}^y,\dots,\theta_{n2}^y, \theta_{12}^{xy},\dots,\theta_{n2}^{xy}). \nonumber
\end{align*}
Then, from Eq.($\ref{eq11}$), the likelihood function can be written as
\begin{equation}\label{eq12}
L_I = L_{I1}\times L_{I2}
\end{equation}
where
\begin{align}\label{eq13}
\begin{split}
L_{I1} &= \prod_{i=1}^n g_1(\theta_{i1}^{x}, \theta_{i1}^{y};\mu_x, \mu_y, \sigma_x^2, \sigma_y^2,\rho)  \\
 &= \prod_{i=1}^n \left(  2\pi\sigma_x\sigma_y(1-\rho^2)^{1/2} \right)^{-1} \exp\bigg\{-\frac{1}{2(1-\rho^2)} \bigg[  \frac{\left(\theta_{i1}^{x}-\mu_x\right)^2}{\sigma^2_x}    \\
     &~~~~~~~~~~~~~~~~~~~+  \frac{ \left( \theta_{i1}^{y}-\mu_y\right)^2 }{\sigma^2_y}
-2\rho\frac{ \left(\theta_{i1}^{x}-\mu_x \right)\left( \theta_{i1}^{y} -\mu_y\right)}{\sigma_x\sigma_y}\bigg]\bigg\}
\end{split}
\end{align}
 and
 {\small
\begin{align}\label{eq14}
L_{I2} &= \prod_{i=1}^n g_2(\theta_{i2}^{x}, \theta_{i2}^{y},\theta_{i2}^{xy}; \gamma_1,\gamma_2,\gamma_3) \nonumber \\
&= \prod_{i=1}^n \frac{  \left(\gamma_1\gamma_2-\gamma_3^2 \right)^{-\frac{\nu}{2}}  }{2 \sqrt[\nu]{\pi}\Gamma(\frac{\nu}{2}) \Gamma(\frac{\nu-1}{2})}   \left(\theta_{i2}^{x}\theta_{i2}^{y} -(\theta_{i2}^{xy})^2 \right)^{\frac{\nu-3}{2}}
       \exp\left\{\frac{-\gamma_1\gamma_2}{2(\gamma_1\gamma_2-\gamma_3^2)} \left( \frac{\theta_{i2}^{x}}{\gamma_1}+\frac{\theta_{i2}^{ y}}{\gamma_2}-\frac{2\gamma_3 \theta_{i2}^{ xy}}{\gamma_1 \gamma_2}\right)\right\}.
\end{align}}


For completeness, the log likelihood and its derivatives with respect to the parameters are shown in  Appendix B. Hence,  we obtain the maximum likelihood estimators $ \hat{\mbox{\boldmath$\tau$}}_{xy} = (\hat{\mu}_x, \hat{\mu}_y, \hat{\sigma}_x^2, \hat{\sigma}_y^2, \hat{\rho}, \hat{\gamma}_1, $ $\hat{\gamma}_2, \hat{\gamma}_3)$ for $\mbox{\boldmath$\tau$}_{xy} = (\mu_x, \mu_y, \sigma_x^2, \sigma_y^2, \rho, \gamma_1,\gamma_2,\gamma_3)$ as
\begin{align}\label{eq24}
 \hat{\mu}_x = \frac{1}{n}\sum_{i=1}^n \theta_{i1}^x, ~~~\hat{\mu}_y = \frac{1}{n}\sum_{i=1}^n \theta_{i1}^y, ~~~~ \hat{\sigma}_x^2 = \frac{1}{n}\sum_{i=1}^n \left( \theta_{i1}^x - \hat{\mu}_x\right)^2,\\ \label{eq25}
 \hat{\sigma}_y^2 = \frac{1}{n}\sum_{i=1}^n \left( \theta_{i1}^y - \hat{\mu}_y\right)^2, ~~~  \hat{\rho} = \frac{\sum_{i=1}^n( \theta_{i1}^y-\hat{\mu}_y)(\theta_{i1}^x-\hat{\mu}_x)}
{\left(\sum_{i=1}^n( \theta_{i1}^y- \hat{\mu}_y)^2\sum_{i=1}^n(\theta_{i1}^x- \hat{\mu}_x)^2\right)^{1/2}},
\end{align}
that is,  $\hat{\rho} = \hat{\sigma}_{xy}/\hat{\sigma}_x\hat{\sigma}_y$ where $\sigma_{xy} = Cov(\Theta_{1}^x, \Theta_{1}^y)$ and hence, the estimator for the covariance is
\begin{equation} \label{eq26} 
 \hat{\sigma}_{xy} = \frac{1}{n}\sum_{i=1}^n( \theta_{i1}^y -\hat{\mu}_y)(\theta_{i1}^x-\hat{\mu}_x),
\end{equation}
and
\begin{equation} \label{eq27} 
\hat{\gamma}_1 = \frac{1}{n\nu}\sum_{i=1}^n  \theta_{i2}^x, ~~~~ \hat{\gamma}_2 = \frac{1}{n\nu}\sum_{i=1}^n  \theta_{i2}^y,~~~~ \hat{\gamma}_3 = \frac{1}{n\nu}\sum_{i=1}^n  \theta_{i2}^{xy}.
\end{equation}

Notice that none of the internal means components  $(\theta_{i1}^x,\theta_{i1}^y)$  and the internal variations components  $(\theta_{i2}^x,\theta_{i2}^y, \theta_{i2}^{xy}) $ is observable. However, each of these unobserved components can be replaced with a suitable realization of the internal distributions. For instance, if we take the standard assumption for interval methodologies that  the internal distributions within the $(X,Y)$, i.e.,  $f_i^x(x_{i}; {\bf \Theta}_i^x)$ and $f_i^y(y_{i}; {\bf \Theta}_i^y)$  are uniformly distributed, then for given interval observations   $X_i=[c_i, d_i]$ and $Y_i=[a_i, b_i]$, realizations of the unobserved internal components are given as
\begin{align}\label{eq28}
\begin{split}
   & \theta_{i1}^x  = (c_i + d_i)/2, ~~~~  \theta_{i1}^y= (a_i + b_i)/2, \\
      \theta_{i2}^x = (d_i- c_i)^2/12,  &~~~~ \theta_{i2}^y= (b_i -a_i)^2/12, ~~~~ \theta_{i2}^{xy}= (d_i- c_i)(b_i -a_i)/12.
\end{split}
\end{align}
Then,  by substituting these realizations into the internal estimators in Eqs.\eqref{eq24}-\eqref{eq27}, we obtain
{\small
\begin{align}\label{eq29}
 &  \hat{\mu}_x = \frac{1}{2n}\sum_{i=1}^n(c_i+d_i), ~~~~~  \hat{\mu}_y = \frac{1}{2n}\sum_{i=1}^n(a_i+b_i), \\\label{eq30}
 & \hat{\sigma}_x^2 = \frac{1}{n}\sum_{i=1}^n[(c_i+d_i)/2 - \hat{\mu}_x]^2,    ~~~~~~~  \hat{\sigma}_y^2 = \frac{1}{n}\sum_{i=1}^n[(a_i+b_i)/2 - \hat{\mu}_y]^2,  \\\label{eq31}
     & \hat{\rho} = \frac{\sum_{i=1}^n[(a_i+b_i)/2-\hat{\mu}_y][(c_i+d_i)/2-\hat{\mu}_x]} { \left( \sum_{i=1}^n[(a_i+b_i)/2- \hat{\mu}_y]^2\sum_{i=1}^n[(c_i+d_i)/2- \hat{\mu}_x]^2\right)^{1/2}}, \\\label{eq32}
    &  \hat{\gamma}_1 = \frac{1}{12n \nu }\sum_{i=1}^n(d_i-c_i)^2,~~  \hat{\gamma}_2 = \frac{1}{12n\nu  }\sum_{i=1}^n(b_i-a_i)^2,~~~ \hat{\gamma}_3 = \frac{1}{12n\nu}\sum_{i=1}^n(b_i-a_i)(d_i-c_i)
\end{align}}
where the  three  estimators in  Eq.\eqref{eq32} ($\hat{\gamma}_1, \hat{\gamma}_2$, $\hat{\gamma}_3$) relate to the \emph{within} interval means and variations given the observations $Y_i = y_{i}$ and $X_i = x_{i}$, $i = 1,\dots,n$, whereas the other estimators given in  Eqs.\eqref{eq29}-\eqref{eq31}   refer to  \emph{between} interval variations.      Thus, for example,  $\hat{\sigma}^2_y$ estimates the variance of the means of the interval values for $Y$; that is, this is the so-called \emph{between} observations variance. Likewise, the estimator $\hat{\sigma}_{xy}=\hat{\rho}\hat{\sigma}_x\hat{\sigma}_y$ estimates the \emph{between} observations covariance of $(X, Y)$.

\subsection{Overall Parameter Estimators}\label{Sec:4.2}
In this section, we obtain the estimators of the  overall means, variances and covariance of the variables $X, ~Y$.
To do this, we need conditional expectations.
For clarity, let us denote the overall $X$ variable by $W^x$ to distinguish it from the conditional $X_i$ values. Likewise, let $W^y$ be the overall $Y$ variable. [Here, the $X,~Y$ variables are the same as those  in Section 3 with probability density  function $h_{X,Y}(x,y; \mbox{\boldmath$\delta$})$.] Then, we have   sets of values of $W^x$ in $X_i = x_{i}$, and   sets of $W^y$ in $Y_i = y_{i}$, with conditional distributions $f_{W^x}(w^x|x_{i})$ with $w^x\in x_{i}$ and  $f_{W^y}(w^y|y_{i})$ with $w^y\in y_{i}$, $i = 1,\dots, n$, and joint conditional  distribution $f_{W^x,W^y}(w^x,w^y|x_{i},y_{i})$ for $(w^x,w^y) \in (x_{i},y_{i})$.

For the within interval variable $X_i$, the mean was taken to be $\Theta_{i1}^x$, so that
\begin{equation}  \label{eq33}
\Theta_{i1}^x = E_{W^x}(W^x|X_i=x_{i}).
\end{equation}
Hence,  we have
\begin{equation}  \label{eq34}
E_{W^x}(W^x) = E_{X_i}[E_{W^x}(W^x|X_i=x_{i})] = E_{X_i}(\Theta_{i1}^x).
\end{equation}
However,  ${\bf \Theta}_{i1} = (\Theta_{i1}^x, \Theta_{i1}^y)$  follows a bivariate normal distribution $N_2(\mu_x, \mu_y, \sigma_x^2, \sigma_y^2, \rho)$ (see Section 4.1). Hence, $\Theta_{i1}^x$ and  $\Theta_{i1}^y$ follow  a normal distribution $N(\mu_x, \sigma_x^2)$ and  $N(\mu_y, \sigma_y^2)$, respectively. Therefore, the overall means of  $W^x$ and $W^y$ are given by
\begin{equation} \label{eq35} 
E_{W^x}(W^x) = \mu_x, ~~~~~~~~E_{W^y}(W^y) = \mu_y.
\end{equation}

To calculate the overall variances, we first recognize that the internal within observation variances given the observations $X_i = x_{i}$ and $Y_i = y_{i}$ were set to be $\Theta_{i2}^x$ and $\Theta_{i2}^y$, respectively. Therefore,

\begin{equation} \label{eq37} 
\Theta_{i2}^x = Var_{W^x}(W^x|X_i=x_{i}),
\end{equation}
for each $i = 1,\dots, n$. Then, we have
\begin{align} \label{eq38}
Var_{W^x}(W^x) &= E_{X_i}[Var_{W^x}(W^x|X_i=x_{i})] + Var_{X_i}[E_{W^x}(W^x|X_i=x_{i})] \nonumber \\
&= E_{X_i}(\Theta_{i2}^x) + Var_{X_i}(\Theta_{i1}^x). 
\end{align}
Now, we have that ${\bf \Theta}_{i2} = (\Theta_{i2}^x, \Theta_{i2}^y, \Theta_{i2}^{xy})$ followed the  bivariate Wishart distribution of Eq.($\ref{eq10}$) and so we can calculate the first term in Eq.($\ref{eq38}$),  $E_{X_i}(\Theta_{i2}^x)$. We also have $\Theta_{i1}^x$ distributed as a normal distribution $N(\mu_x, \sigma_x^2)$. Hence, Eq.($\ref{eq38}$) becomes
\begin{equation} \label{eq39} 
Var_{W^x}(W^x) =   \nu  \gamma_1 + \sigma^2_x
\end{equation}
where $\nu $ is the degree freedom of the bivariate Wishart distribution.
Similarly, we can show that
\begin{equation}  \label{eq40}
Var_{W^y}(W^y) =  \nu \gamma_2 + \sigma^2_y.
\end{equation}

To obtain the corresponding expression for the covariance, we can show that
\begin{align} \label{eq41}
Cov_{W^x,W^y}(W^x,W^y) &= E_{X_i,Y_i}[Cov_{W^x,W^y}(W^x,W^y|X_i=x_{i},Y_i=y_{i},)] \nonumber \\
&~~~~+ Cov_{X_i,Y_i}[E_{W^x}(W^x|X_i=x_{i}), E_{W^y}(W^y|Y_i=y_{i})]\nonumber \\
&= E_{X_i,Y_i}(\Theta_{i2}^{xy}) + Cov_{X_i,Y_i}(\Theta_{i1}^x,\Theta_{i1}^y). 
\end{align}

The first term of the right-hand side  in Eq.($\ref{eq41}$) is the covariance  of  the bivariate Wishart distribution given in  Eq.($\ref{eq10}$), from which we can show that the internal expectation $E_{X_i,Y_i}(\Theta_{i2}^{xy}) = \nu \gamma_3$. The second term in Eq.($\ref{eq41}$) corresponds to the covariance of the two internal means $\Theta_{i1}^x,\Theta_{i1}^y$. Since it is assumed that the joint distribution of $(\Theta_{i1}^x,\Theta_{i1}^y)$ follows the bivariate normal distribution $N_2(\mu_x, \mu_y, \sigma_x^2, \sigma_y^2, \rho)$, it follows that this covariance is $\rho\sigma_x\sigma_y.$ Hence, substituting into Eq.($\ref{eq41}$), we obtain
\begin{equation} \label{eq42}
Cov_{W^x,W^y}(W^x,W^y) =  \nu \gamma_3 + \rho\sigma_x\sigma_y.
\end{equation}

All these overall moments are expressed in terms of the parameters contained in $\mbox{\boldmath$\tau$}$. Hence, the overall maximum likelihood estimators are readily found by substituting the relevant values from Eqs.($\ref{eq29}$)-($\ref{eq32}$). That is,
\begin{equation} \label{eq43} 
 \widehat{E(W^x)} = \hat{\mu}_x = \frac{1}{2n}\sum_{i=1}^n(c_i+d_i), ~~~~ \widehat{E(W^y)} = \hat{\mu}_y =  \frac{1}{2n}\sum_{i=1}^n(a_i+b_i),
\end{equation}
\begin{equation}\label{eq45.1}
   \widehat{Var(W^x)}=  \nu  \hat{\gamma}_1 + \hat{\sigma}^2_x= \frac{1}{12n}\sum_{i=1}^n(d_i-c_i)^2 + \frac{1}{n}\sum_{i=1}^n[(c_i+d_i)/2 - \hat{\mu}_x]^2,
\end{equation}
 \begin{equation}\label{eq46.1}
\widehat{Var(W^y)} = \nu  \hat{\gamma}_2 + \hat{\sigma}^2_y= \frac{1}{12n}\sum_{i=1}^n(b_i-a_i)^2 + \frac{1}{n}\sum_{i=1}^n[(a_i+b_i)/2 - \hat{\mu}_y]^2,
\end{equation}
and
\begin{align} \label{eq47.1}
\begin{split}
\widehat{Cov(W^x,W^y)} &=  \nu \hat{\gamma}_3 + \hat{\rho} \hat{\sigma}_x\hat{\sigma}_y=  \frac{1}{12n}\sum_{i=1}^n(b_i-a_i)(d_i-c_i) \\
&~~~~~~~~~~~~~~~~~~~~~~~~~~~~+  \frac{1}{n}\sum_{i=1}^n[(a_i+b_i)/2-\hat{\mu}_y][(c_i+d_i)/2-\hat{\mu}_x].
\end{split}
\end{align}
In Eqs.($\ref{eq43}$)-($\ref{eq47.1}$), it is implicit that these are estimators with respect to the overall variables in that $E_{W^x}(W^x)$, e.g., is written as $E(W^x)$.

\begin{remark}
Since  $E_{X_i}\left( \Theta_{i2}^x\right)= \nu \gamma_1$,  we have  $E\left(\widehat{Var(W^x)} \right)= \frac{\nu}{12}  \gamma_1 +  \sigma^2_x$. Therefore, $\widehat{Var(W^x)} $  would be  unbiased for $ \gamma_1 +  \sigma^2_x $ if $\nu=12$.   This is also true for $\widehat{Var(W^y)}$ and $\widehat{Cov(W^x,W^y)} $ in Eqs.\eqref{eq46.1}-\eqref{eq47.1}.  Hence, to obtain unbiased estimators, we  rewrite the overall estimators by replacing the divisor  $12$ with the degree of freedom $\nu$  as follows
\end{remark}
\begin{equation}\label{eq45.2}
  S^2_X= \widehat{Var(W^x)}=   \frac{1}{\nu  n}\sum_{i=1}^n(d_i-c_i)^2 + \frac{1}{n}\sum_{i=1}^n[(c_i+d_i)/2 - \hat{\mu}_x]^2,
\end{equation}
 \begin{equation}\label{eq46.2}
S^2_Y= \widehat{Var(W^y)} =  \frac{1}{\nu  n}\sum_{i=1}^n(b_i-a_i)^2 + \frac{1}{n}\sum_{i=1}^n[(a_i+b_i)/2 - \hat{\mu}_y]^2,
\end{equation}
and
\begin{align} \label{eq47.2}
\begin{split}
 S^2_{XY}=\widehat{Cov(W^x,W^y)} &=   \frac{1}{\nu n}\sum_{i=1}^n(b_i-a_i)(d_i-c_i) \\
& ~~~~~~~~~~~~~~~~~~+
  \frac{1}{n}\sum_{i=1}^n[(a_i+b_i)/2-\hat{\mu}_y][(c_i+d_i)/2-\hat{\mu}_x].
\end{split}
\end{align}

The overall mean $\mu_x$ (and similarly for $\mu_y$) is estimated by the average of the interval midpoints. Thus, regardless of the actual length of an interval, the sample mean is unchanged. This result was first obtained empirically by Bertrand and Goupil (2000) using an empirical distribution approach.

Bertrand and Goupil (2000) also obtained an expression for the sample variance. Later, Billard (2008) showed that their sample variance was the sum of a  \emph{Within Observation} variation and a \emph{Between Observation} variation. In the overall variance $Var(W^x)$ of Eq.($\ref{eq39}$), the two terms $\gamma_1$ and $\sigma_x^2$  correspond, respectively, to the \emph{Within Observations} variation and the \emph{Between Observations} variation.  [Clearly, the same applies to $Var(W^y)$ of Eq.($\ref{eq40}$).] The same phenomenon applies to the $Cov(X, Y)$ with the first term $\gamma_3$ in Eq.($\ref{eq42}$) corresponding to the \emph{Within Observation} covariation and the second term $\rho\sigma_x\sigma_y = \sigma_{xy}$ corresponding to the \emph{Between Observation} covariation.
 This result was initially obtained as a moment estimator of the covariance function in Billard (2008).

\begin{remark}
 With some algebra, it can  be shown that (when $\nu=12$) the overall maximum likelihood estimators of the variances and the covariance in Eqs.\eqref{eq45.1}-\eqref{eq47.1}  are  identical to their corresponding empirical  versions proposed  by Billard (2008) given in Eqs.\eqref{eq2}-\eqref{eq3}. That is,
 \begin{equation} \label{eq45} 
S^2_{X}=\widehat{Var(W^x)} = \frac{1}{3n}\sum_{i=1}^n [(c_i-\hat{\mu}_x)^2 + (c_i-\hat{\mu}_x)(d_i-\hat{\mu}_x) + (d_i-\hat{\mu}_x)^2],
\end{equation}
\begin{equation} \label{eq46} 
S^2_{Y}=\widehat{Var(W^y)} = \frac{1}{3n}\sum_{i=1}^n [(a_i-\hat{\mu}_y)^2 + (a_i-\hat{\mu}_y)(b_i-\hat{\mu}_y) + (b_i-\hat{\mu}_y)^2],
\end{equation}
and
\begin{align} \label{eq47}
S_{XY}=\widehat{Cov(W^x,W^y)} &= \frac{1}{6n}\sum_{i=1}^n [2(a_i-\hat{\mu}_y)(c_i-\hat{\mu}_x) + (a_i-\hat{\mu}_y)(d_i-\hat{\mu}_x) \nonumber \\
&~~~~~~~~~~+ (b_i-\hat{\mu}_y)(c_i-\hat{\mu}_x) + 2(b_i-\hat{\mu}_y)(d_i-\hat{\mu}_x)].
\end{align}
\end{remark}

\begin{remark}
When the  data are classically-valued, i.e., when $x_{i} = [a_i, a_i] = a_i$ and $y_{i} = [c_i, c_i] = c_i$, it is easy to show that the Within Observation variations become zero, while the Between Observation variations are unchanged since they are based on the interval midpoints. In this case, all the results in  Eqs.($\ref{eq43}$)-($\ref{eq47.1}$)  reduce to their classical counterparts, as they should, thus indirectly verifying and corroborating the veracity of the derivations herein.
\end{remark}

\section{Asymptotic Properties of the MLEs}\label{Sec:asympto}
In this section, we study  the consistency  and asymptotic normality of the  maximum likelihood estimators (MLE) estimators proposed in Section 4.

\begin{theorem}  (Consistency) Under some regularity conditions on the  underlying  family of   distributions, the MLE estimators are consistent, i.e.,   (where $\overset{p}{\longrightarrow}$ denotes  convergence in probability)

(a) ~ $\widehat{\mu}_x \overset{p}{\longrightarrow} \mu_x $, and ~ $\widehat{\mu}_y \overset{p}{\longrightarrow} \mu_y $;

(b) ~ $\widehat{\sigma}_x^2 \overset{p}{\longrightarrow} \sigma_x^2$,  ~$\widehat{\sigma}_y^2 \overset{p}{\longrightarrow} \sigma_y^2$, and ~$\widehat{\rho} \overset{p}{\longrightarrow} \rho $;

(c) ~ $\widehat{\gamma}_1 \overset{p}{\longrightarrow}  \gamma_1 $, ~ $\widehat{\gamma}_2 \overset{p}{\longrightarrow}  \gamma_2 $, and ~ $\widehat{\gamma}_3 \overset{p}{\longrightarrow}  \gamma_3 $.
\end{theorem}

Since $(\Theta_{i1}^x, \Theta_{i1}^y) \sim N_2(\mu_x, \mu_y, \sigma_x^2, \sigma_y^2, \rho)$,  and    $ \widehat{\mu}_x = \frac{1}{n}\sum_{i=1}^n \theta_{i1}^x$ and   $\widehat{\mu}_y = \frac{1}{n}\sum_{i=1}^n \theta_{i1}^y$,     we have   $\sqrt{  n} \left( \widehat{\mu}_x  -\mu_x     \right) \sim   N\left(0, \sigma^2_x   \right)  $  and  $\sqrt{  n} \left( \widehat{\mu}_y -\mu_y  \right) \sim   N\left(0, \sigma^2_y   \right)$.
  However, the  central limit theorem  is used to obtain  the asymptotic distributions of the other MLE estimators given in Eqs.\eqref{eq24}-\eqref{eq27}. They are summarized in the following theorem.

\begin{theorem}  (Asymptotic Normality) By the central limit theorem,    we have

(a)  $\sqrt{  n} \left( \widehat{\sigma}_x^2 -\sigma^2_x  \right)\overset{D}{\longrightarrow}  N\left(0, 2 \sigma^4_x   \right)  $,  ~$\sqrt{  n} \left( \widehat{\sigma}_y^2 -\sigma^2_y  \right)\overset{D}{\longrightarrow}  N\left(0, 2 \sigma^4_y   \right)  $,  and ~

 ~~~ $\sqrt{n} \left( \widehat{\sigma}_{xy}  -\sigma_{xy}  \right)\overset{D}{\longrightarrow}  N\left(0,  (1+\rho^2) \sigma^2_x \sigma^2_y   \right)  $;

(b) $\sqrt{\nu n} \left( \widehat{\gamma}_1 -\gamma_1  \right)\overset{D}{\longrightarrow}  N\left(0,   2 \gamma_1^2  \right)  $, ~ $\sqrt{ \nu n} \left( \widehat{\gamma}_2 -\gamma_2  \right)\overset{D}{\longrightarrow}  N\left(0,   2 \gamma_2^2 \right)  $, ~ and

   ~~~  $\sqrt{ \nu n} \left( \widehat{\gamma}_3-\gamma_3  \right)\overset{D}{\longrightarrow}  N\left(0,    \gamma_1\gamma_2 + \gamma_3^2  \right)$;

where    $\nu $ is the degree of freedom of the  bivariate Wishart distribution, and   $\overset{D}{\longrightarrow}$ denotes  convergence in distribution. 
\end{theorem}

Now, we can combine these results to obtain the asymptotic distributions of the overall MLE  estimators of the interval variables given in Eqs.\eqref{eq45.2}-\eqref{eq47.2}.  They are indicated  in the following theorem. The proofs  are omitted.

\begin{theorem}  (Asymptotic Normality)

(a) ~  $ S^2_X   \overset{D}{\longrightarrow}  N\left( \gamma_1 + \sigma^2_x ,  ~  \frac{2(\gamma_1^2 + \nu \sigma^4_x)}{\nu n} \right) $;

(b) ~  $ S^2_Y   \overset{D}{\longrightarrow}  N\left( \gamma_2 + \sigma^2_y , ~   \frac{2(\gamma_2^2 +  \nu \sigma^4_y)}{\nu  n} \right) $;

(c) ~  $ S^2_{XY}   \overset{D}{\longrightarrow}  N\left( \gamma_3 + \rho \sigma_x \sigma_y , ~   \frac{ (\gamma_1 \gamma_2 + \gamma_3^2 )+ \nu  (1+\rho^2) \sigma^2_x \sigma^2_y  }{\nu n} \right) $.

\end{theorem}

\section{Extensions and Generalizations}\label{Sec:6}

The results in Section 4 have been derived under an assumption that the within interval observations are uniformly distributed across the given intervals. Other distributions could be used. In those cases, realizations of the interval parameters $\Theta_{i1}^x$, etc., will change. These will give different expressions for the Within Observation quantities, while those for Between Observation terms are unchanged. The principles are the same however as were followed in the above derivations. We illustrate briefly the case where the observations across an interval are distributed according to a triangular distribution (in Section 6.1) or a Pert distribution (in Section 6.2). 

\subsection{Triangular Interval Data}

As before, we maintain the assumption that the interval means are normally distributed and that the internal variations follow a bivariate Wishart distribution.  However, instead of observations  being uniformly spread across intervals, suppose now
 we assume values within random  intervals $X_i=[c_i, d_i]$ and $Y_i=[a_i, b_i]$, i.e.,    $f_i^x(x_{i}; {\bf \Theta}_i^x)$ and $f_i^y(y_{i}; {\bf \Theta}_i^y)$ of Eq.($\ref{eq5}$), are triangularly distributed.
This internal distribution represents intervals for which aggregated observations within an interval are clustered more around a  central value (say, $\zeta$) of the interval rather than being uniformly spread across the interval. In the context of symbolic interval data, it is not unreasonable that $\zeta = (a + b)/2$; thus,  we illustrate the theory for this case.

The within observation random variables  ${\bf \Theta}_{i1} = (\Theta_{i1}^x, \Theta_{i1}^y)$ and ${\bf \Theta}_{i2} = (\Theta_{i2}^x, \Theta_{i2}^y, \Theta_{i2}^{xy})$ now take realizations $\theta_{i1}^y = (a_i + b_i)/2$, $\theta_{i2}^x = (c_i + d_i)/2$, $\theta_{i2}^y = (b_i - a_i)^2/24$, $\theta_{i2}^x = (d_i - c_i)^2/24$, and $\theta_{i2}^{xy} = (b_i - a_i)(d_i - c_i)/24$, respectively. These are then substituted into the internal estimators in Eqs.\eqref{eq24}-\eqref{eq27}. After the necessary derivations, we can show that the maximum likelihood estimators $(\hat{\mu}_x, \hat{\mu}_y, \hat{\sigma}_x^2, \hat{\sigma}_y^2, \hat{\rho})$ for $(\mu_x, \mu_y, \sigma_x^2, \sigma_y^2, \rho)$  correspond to those given in    Eqs.($\ref{eq29}$)-($\ref{eq31}$); while for  $(\gamma_1,\gamma_2,\gamma_3)$, the maximum likelihood estimators are, respectively,
\begin{equation*} 
\hat{\gamma}_1 = \frac{1}{n}\sum_{i=1}^n(d_i-c_i)^2/24,~~
\hat{\gamma}_2 = \frac{1}{n}\sum_{i=1}^n(b_i-a_i)^2/24,~~
\hat{\gamma}_3 = \frac{1}{n}\sum_{i=1}^n(b_i-a_i)(d_i-c_i)/24.
\end{equation*}

Therefore, by substituting these into the conditional moments,  Eq.($\ref{eq35}$) and  Eqs.($\ref{eq39}$)-($\ref{eq42}$), we obtain the overall maximum likelihood estimators to be
\begin{align*} 
&\widehat{E(W^x)} = \hat{\mu}_x = \frac{1}{2n}\sum_{i=1}^n(c_i+d_i), ~~~~~~~~ \widehat{E(W^y)} = \hat{\mu}_y =  \frac{1}{2n}\sum_{i=1}^n(a_i+b_i),\\ 
&\widehat{Var(W^x)} = \frac{1}{24n}\sum_{i=1}^n [7(c_i-\hat{\mu}_x)^2 + 10(c_i-\hat{\mu}_x)(d_i-\hat{\mu}_x) + 7(d_i-\hat{\mu}_x)^2],\\ 
&\widehat{Var(W^y)} = \frac{1}{24n}\sum_{i=1}^n [7(a_i-\hat{\mu}_y)^2 + 10(a_i-\hat{\mu}_y)(b_i-\hat{\mu}_y) + 7(b_i-\hat{\mu}_y)^2],
\end{align*}
and
\begin{align*}  
\widehat{Cov(W^x,W^y)} &= \frac{1}{24n}\sum_{i=1}^n [7(a_i-\hat{\mu}_y)(c_i-\hat{\mu}_x) + 5(a_i-\hat{\mu}_y)(d_i-\hat{\mu}_x) \nonumber \\
&~~~~~~~~~~~~~~~~~~~~~+ 5(b_i-\hat{\mu}_y)(c_i-\hat{\mu}_x) + 7(b_i-\hat{\mu}_y)(d_i-\hat{\mu}_x)].
\end{align*}
We observe that these maximum likelihood estimators are the same as the empirical  moment estimators with a triangular internal distribution proposed by  Billard (2008).

\subsection{ Pert Interval Data}
 Another possibility  for the internal  distribution  is  to assume the  observations within an interval follow a Pert distribution.  The Pert distribution also known as a Beta-Pert distribution, is a non-uniform bounded support distribution that  is  very  flexible and robust with respect to most types of skewed distributions on the given range of $[a, b]$. Like the Triangular distribution, the Pert distribution also uses the most likely value and is designed to produce a distribution that accurately reflects the true distribution.
This  distribution was introduced by Malcolm et al. (1959) and Clark (1962). Let $Y\sim P(a, b, m_y)$, then the probability density function of $Y$ is defined as
\begin{align*}
 f(y)= \frac{(y-a)^{\alpha_1-1}(b-y)^{\alpha_2-1}}{ B(\alpha_1, \alpha_2)(b-a)^{\alpha_1+\alpha_2-1}}, ~~~~~~ a<y<b,
\end{align*}
where $\alpha_1=6(\mu_y-a)(b-a)$, $\alpha_2=6(b-\mu_y)(b-a)$, with mean value $\mu_y =(a+4m_y+b)/6$, such that $m_y$ is the most likely value of the  random variable $Y$, and $B(\alpha_1, \alpha_2)$ is the Beta function.

 The realizations of  the within observation random parameters  ${\bf \Theta}_{i1} = (\Theta_{i1}^x, \Theta_{i1}^y)$ and ${\bf \Theta}_{i2} = (\Theta_{i2}^x, \Theta_{i2}^y, \Theta_{i2}^{xy})$ now become  $\theta_{i1}^y = (a_i + 4m_{y_i}+ b_i)/6$, $\theta_{i2}^x = (c_i +4m_{x_i}+ d_i)/6$, $\theta_{i2}^y = (\mu_{y_i}-a_i)(b_i - \mu_{y_i})/7$, $\theta_{i2}^x = (\mu_{x_i}-c_i)(d_i - \mu_{x_i})/7$, and $\theta_{i2}^{xy} = \big[ (\mu_{x_i}-c_i)(b_i - \mu_{y_i}) +  (\mu_{y_i}-a_i)(d_i - \mu_{x_i}) \big]  /14$, respectively.
 Then,   the maximum likelihood estimators $ \hat{\mbox{\boldmath$\tau$}}_{xy} = (\hat{\mu}_x, \hat{\mu}_y, \hat{\sigma}_x^2, \hat{\sigma}_y^2, \hat{\rho}, \hat{\gamma}_1,\hat{\gamma}_2,\hat{\gamma}_3)$ for $ \mbox{\boldmath$\tau$}_{xy} = (\mu_x, \mu_y, \sigma_x^2, \sigma_y^2, \rho, \gamma_1,\gamma_2,\gamma_3)$  are obtained as follows
\begin{equation*}
\hat{\mu}_x =  \frac{1}{n}\sum_{i=1}^n \mu_{x_i}=\frac{1}{6n}\sum_{i=1}^n(c_i+4m_{x_i}+d_i), ~~~~\hat{\mu}_y =  \frac{1}{n}\sum_{i=1}^n \mu_{y_i}=\frac{1}{6n}\sum_{i=1}^n(a_i+ 4m_{y_i}+b_i),
\end{equation*}
\begin{equation*}
\hat{\sigma}_x^2 = \frac{1}{n}\sum_{i=1}^n[(c_i+ 4m_{x_i} +d_i)/6 - \hat{\mu}_x]^2, ~~~~ \hat{\sigma}_y^2 = \frac{1}{n}\sum_{i=1}^n[(a_i+4m_{y_i}+b_i)/6 - \hat{\mu}_y]^2,
\end{equation*}
\begin{equation*} 
 \hat{\sigma}_{xy} = \frac{1}{n}\sum_{i=1}^n[(a_i+4m_{y_i}+b_i)/6-\hat{\mu}_y][(c_i+4m_{x_i}+d_i)/6-\hat{\mu}_x],
\end{equation*}
\begin{equation*}
\hat{\gamma}_1 = \frac{1}{n}\sum_{i=1}^n(\mu_{x_i}-c_i)(d_i - \mu_{x_i})/7, ~~~~~~~~~ \hat{\gamma}_2 = \frac{1}{n}\sum_{i=1}^n(\mu_{y_i}-a_i)(b_i - \mu_{y_i})/7,
\end{equation*}
and
\begin{equation*}
\hat{\gamma}_3 = \frac{1}{n}\sum_{i=1}^n\big[ (\mu_{x_i}-c_i)(b_i - \mu_{y_i}) +  (\mu_{y_i}-a_i)(d_i - \mu_{x_i}) \big]  /14.
\end{equation*}

Thence, substituting these estimators  into the conditional moments,  Eq.($\ref{eq35}$) and  Eqs.($\ref{eq39}$)-($\ref{eq42}$), we obtain the overall maximum likelihood estimators  for internal  Pert distributed data as
\begin{equation*}
 \widehat{E(W^x)} = \hat{\mu}_x = \frac{1}{6n}\sum_{i=1}^n(c_i+4m_{x_i}+d_i), ~~~~ \widehat{E(W^y)} = \hat{\mu}_y =  \frac{1}{6n}\sum_{i=1}^n(a_i+ 4m_{y_i}+b_i),
\end{equation*}
\begin{align*}
& \widehat{Var(W^x)} =  \frac{1}{7  n }\sum_{i=1}^{n} (\mu_{x_i}-c_{i})(d_{i}-\mu_{x_i})
  +  \frac{1}{ n }\sum_{t=1}^{n}  \left((c_{i} + 4m_{x_i} + d_{i})/6 -\hat{\mu}_x\right)^2,\\ 
   & \widehat{Var(W^y)} = \frac{1}{7  n }\sum_{i=1}^{n} (\mu_{y_i}-a_{i})(b_{i}-\mu_{y_i})
  +  \frac{1}{ n }\sum_{t=1}^{n}  \left((a_{i} + 4m_{y_i} + b_{i})/6 -\hat{\mu}_y\right)^2,
\end{align*}
and
\begin{align*}
\begin{split}
\widehat{Cov(W^x,W^y)} &=   \frac{1}{14  n }\sum_{i=1}^{n} [(\mu_{x_i}-c_i)(b_{i}-\mu_{y_i})+ (\mu_{y_i}-a_{i})(d_{i}-\mu_{x_i})]  \\
  ~~~~~~~ &~~~~~  +  \frac{1}{ n }\sum_{t=1}^{n} \left((c_i + 4m_{x_i} + d_i)/6 -\hat{\mu}_x \right) \left((a_{i} + 4m_{y_i} + b_{i})/6 -\hat{\mu}_y\right).
\end{split}
\end{align*}
In practice,  the most likely value $m_x$ and $m_y$  of the data may not be available; then, it is not unreasonable to assume that the mode of the interval-valued variables $X$ and $Y$ are,  respectively, $m_x=(c+d)/2$ and $m_y=(a+b)/2$.

\section{Some Data}\label{Sec:7}
The foregoing theory is applied to simulated and real data sets. In Section 7.1, data are simulated for two different sets of the parameters with different sample sizes.
 Then, in Section 7.2, a faces data set of interval-valued observations is considered in which  the variance-covariance matrix is calculated and used in a principal component analysis.

\subsection{Simulations}
 To conduct a simulation study, random samples of $n$ two-dimensional intervals $(X, Y)$, were generated. For each sample, the overall  sample  estimators and their corresponding asymptotic variances were  calculated. This is  repeated $B$ times; and the average of each of  ten  descriptive statistics was  calculated.

Centers of the intervals were simulated according to a bivariate normal distribution with means ($\mu_x, \mu_y$),  standard deviations ($\sigma_x, \sigma_y$), and covariance $\sigma_{xy}$. The internal variates for each interval were simulated by a bivariate Wishart distribution with parameters  $\gamma_1$, $\gamma_2$, and $\gamma_3$.   
This gives  the  values ($r_1, r_2$),  which are the marginal ranges of the intervals.
Then, the $X = [c, d]$ is obtained from $X = [x - \sqrt{r_1}/2,~ x +\sqrt{ r_1}/2]$ and likewise $Y = [y - \sqrt{r_2}/2, ~y + \sqrt{r_2}/2]$.

 We take  two  sets of $\boldsymbol\tau= (\mu_x, \mu_y, \sigma_x^2, \sigma_y^2, \sigma_{xy}, \gamma_1, \gamma_2, \gamma_3)$ values,  for each  sample size $n = 50, 100, $ $500, 1000$ and $B = 1000$ iterations. The results were consistent across all parameter sets and all sample sizes. Table 1 reports  the simulation results obtained for  all sample sizes  when $ \boldsymbol\tau =(\mu_x=1,\mu_y=5,\sigma^2_x= 4, \sigma^2_y=3, \sigma_{xy}=2, \gamma_1=7,\gamma_2=5,\gamma_3=-2)$.    The corresponding  overall estimate values when observations are simulated with $ \boldsymbol\tau =(\mu_x=-2,\mu_y=3,\sigma^2_x= 1.5, \sigma^2_y=2.5, \sigma_{xy}=-1.75, \gamma_1=1.25,\gamma_2=2.5,\gamma_3=-1.75)$   are shown in Table 2.   Also   the standard deviations of each estimated value from the $B = 1000$ iterations are given in  parenthesis.  Note that $\widehat{g(\boldsymbol\tau )}$ in Table 1 and Table 2  represents  the vector of the  estimators of interest  following  with their estimated asymptotic variance components  in an every other manner,   which  is   defined as $\widehat{g(\boldsymbol\tau )}= \big(\widehat{\mu}_x, ~n Var(\widehat{\mu}_x), ~\widehat{\mu}_y, ~n Var(\widehat{\mu}_y),~ S^2_X, n Var(S^2_X),~ S^2_Y,   ~ n Var(S^2_Y),  ~ S_{XY},~ n Var(S_{XY}) \big)$  with the corresponding parameter function
 $ g(\boldsymbol\tau )= \big( \mu_x,~ \sigma^2_x,~ \mu_y, ~\sigma^2_y,   ~ (\gamma_1 + \sigma^2_x), ~ 2/\nu(\gamma_1^2 + \nu \sigma_x^4),~   (\gamma_2+\sigma^2_y),~ 2/\nu(\gamma_2^2  +  \nu \sigma_y^4),~ (\gamma_3+ \sigma_{xy}),~  (\gamma_1 \gamma_2 + \gamma_3^2)/\nu+  \sigma^2_x \sigma^2_y(1+\rho^2)  \big)$.

 The  simulation results are good and compatible with the asymptotic results given in Section \ref{Sec:asympto}. In all cases, it is seen that the estimated statistics are very close to the theoretical parameter values.  It is also observed that as the sample size $n$ increases,  values for the respective standard deviations decrease without exception.

\begin{table}[!htbp]
\caption{ Simulation results for estimators of the parameters of interest and their estimated  asymptotic variances provided  in    $\widehat{g(\boldsymbol\tau )}$  with   $ \boldsymbol\tau =(\mu_x=1,\mu_y=5,\sigma^2_x= 4, \sigma^2_y=3, \sigma_{xy}=2, \gamma_1=7,\gamma_2=5,\gamma_3=-2)$, $B = 1000$} 
\centering
 \scalebox{.90}{%
\begin{tabular}{|c|c|cccc|}\hline
$g(\boldsymbol\tau )$ &   $\widehat{g(\boldsymbol\tau )}$  &  $n=50$    &  $n=100$ & $ n=500$             &  $n=1000$           \\ \hline
  1    & $\widehat{\mu}_x$    & 1.000(0.023)& 1.001(0.016)& 1.000(0.007)& 1.000(0.005)
   \\
     4   &  $n Var(\widehat{\mu}_x)$    &4.008(0.806)& 3.995(0.579)& 3.985(0.249)& 4.011(0.177)
   \\
5   & $\widehat{\mu}_y$      &4.999(0.020)& 5.000(0.014)& 5.000(0.007)& 5.000(0.004)
   \\ 
   3   & $n Var(\widehat{\mu}_y)$      & 3.003(0.614)& 3.008(0.410)& 2.995(0.186)& 3.000(0.136)
   \\
 7.25  &  $S^2_X$   &7.220(0.066)& 7.224(0.047)& 7.224(0.021)& 7.223(0.015)
 \\
 33.76 & $n Var(S^2_X)$   & 33.667(6.945)& 33.623(4.958)& 33.538(2.105)& 33.572(1.590)
 \\ 
4.25 & $S^2_Y $   & 4.228(0.048)& 4.230(0.035)& 4.231(0.015)& 4.230(0.011)
  \\
18.26   &  $n Var(S^2_Y) $  & 18.064(3.693)& 18.236(2.596)& 18.175(1.167)& 18.212(0.818)
  \\ 
0  &  $S_{XY}$  &0.011(0.045)& 0.002(0.035)& 0.002(0.015)& 0.002(0.011)
     \\
16.67  & $n Var(S_{XY}$) &  16.677(3.354)& 16.716(2.408) & 16.592(1.090) & 16.598(0.752)
     \\  \hline
\end{tabular}
}
\end{table}

\begin{table}[!htbp]
\caption{ Simulation results for estimators of the parameters of interest and their estimated  asymptotic variances provided  in    $\widehat{g(\boldsymbol\tau )}$ with  $ \boldsymbol\tau =(\mu_x=-2,\mu_y=3,\sigma^2_x= 1.5, \sigma^2_y=2.5, \sigma_{xy}=-1.75, \gamma_1=1.25,\gamma_2=2.5,\gamma_3=-1.75)$, $B = 1000$} 
\centering
 \scalebox{.90}{%
\begin{tabular}{|c|c|cccc|}\hline
$g(\boldsymbol\tau )$ &   $\widehat{g(\boldsymbol\tau )}$  &  $n=50$    &  $n=100$ & $ n=500$             &  $n=1000$           \\ \hline
  -2    & $\widehat{\mu}_x$    &-2.000(0.014)& -2.000(0.010)& -2.000(0.004)& -2.000(0.003)
   \\
     1.5  & $n Var(\widehat{\mu}_x)$    &1.495(0.299)& 1.481(0.205)& 1.502(0.096)& 1.496(0.067)
   \\
3  & $\widehat{\mu}_y$      &3.000(0.018)& 3.000(0.013)& 3.000(0.006)& 3.000(0.004)
   \\ 
  2.5  & $n Var(\widehat{\mu}_y)$      &2.493(0.502)& 2.474(0.352)& 2.507(0.158)& 2.495(0.112)
   \\
2.75  &  $S^2_X$   &2.738(0.024)& 2.740(0.018)& 2.740(0.008)& 2.740(0.006)
 \\
4.76 & $n Var(S^2_X)$   &4.701(0.970)& 4.747(0.715)& 4.726(0.295)& 4.733(0.217)
 \\ 
5 & $S^2_Y $   &4.982(0.041)& 4.983(0.031)& 4.983(0.013)& 4.984(0.010)
  \\
13.54   &  $n Var(S^2_Y) $  &13.430(2.678)& 13.470(1.880)& 13.468(0.872)& 13.461(0.600)
  \\ 
-3.5  &  $S_{XY}$  &-3.503(0.030)& -3.505(0.023)& -3.504(0.010)& -3.505(0.007)
     \\
7.328  & $n Var(S_{XY} $   )& 7.263(1.486)& 7.288(1.074)& 7.282(0.462)& 7.281(0.328)
     \\  \hline
\end{tabular}
}
\end{table}

\subsection{Real Data}
\begin{figure}[!htbp]
  \centering
\includegraphics[height=7cm,width=11cm]{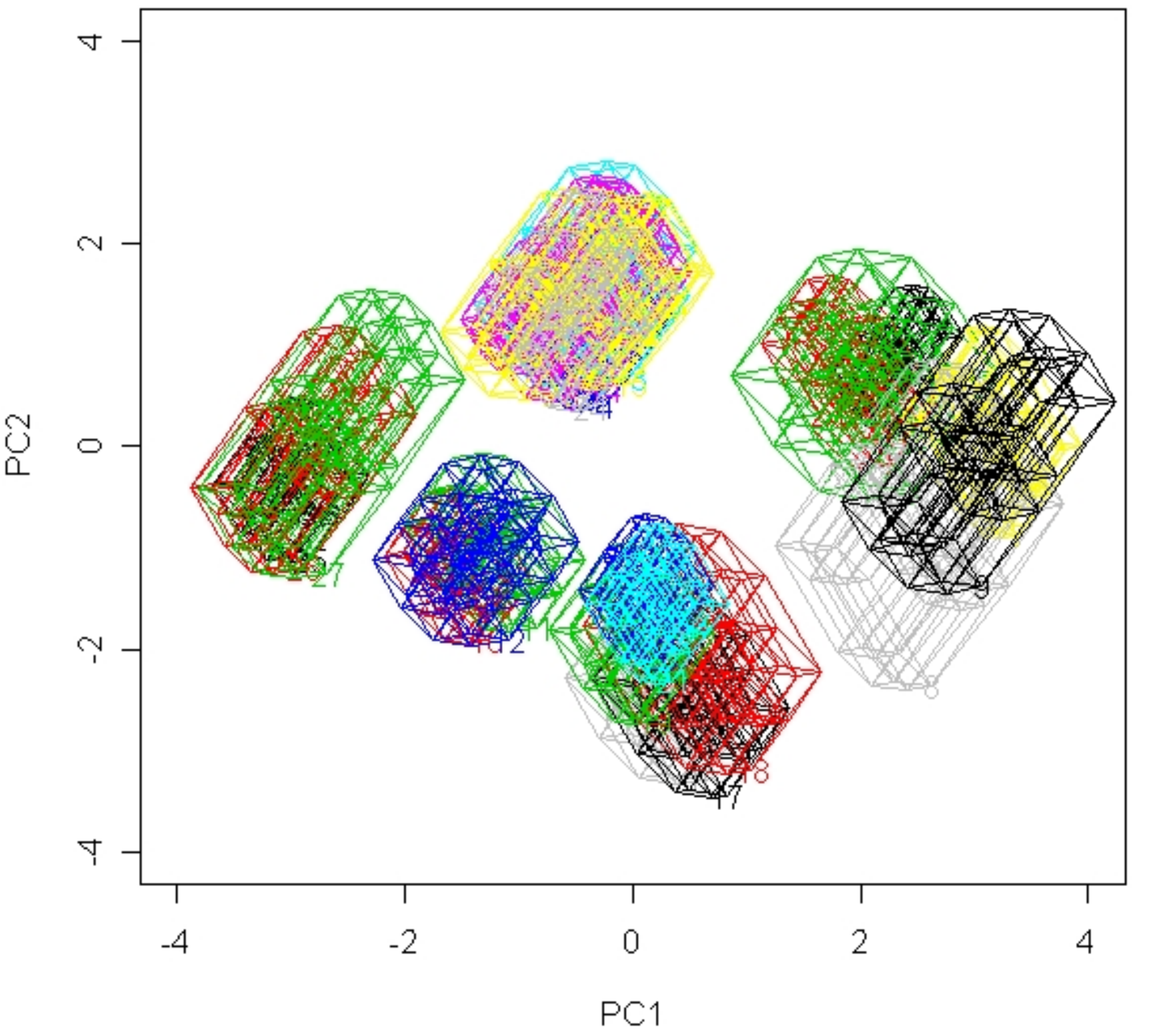}
\caption{Faces Data -  $PC_1~ \times PC_2$ Polytopes} \label{fig1}
\end{figure}

We show the effect of using the covariance function derived herein on a principal component analysis of the Leroy et al. (1996) faces data set, available in Douzal--Chouakria et al. (2011). The data are interval-valued (as a result of aggregation), with detailed descriptions  found in Douzal--Chouakria et al. (2011). There are six variables (eye span, distance between eyes, distance from outer right (respectively, left) eye to the upper middle lip, and the length from the middle lip to the left (respectively, right) mouth for each of  twenty-seven faces. The resulting plots of the first and second principal component analysis using the covariances from Eq.($\ref{eq3}$) and the polytope method of Le-Rademacher and Billard (2012) are shown in Figure $\ref{fig1}$.

\begin{figure}[!htbp]
  \centering
  \begin{tabular}{cc}
\includegraphics[height=4cm,width=5.5cm]{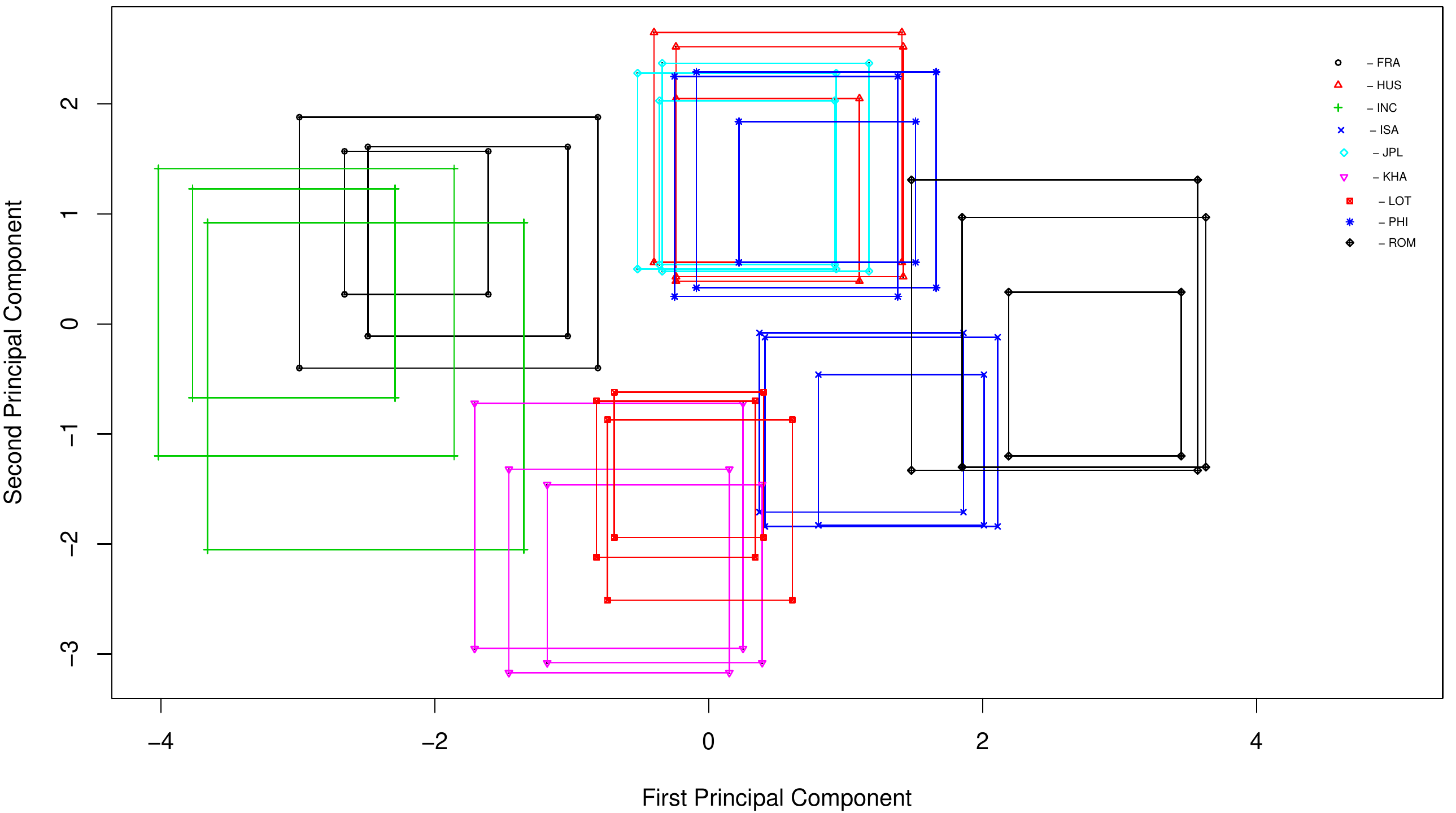}
& \includegraphics[height=4cm,width=5.5cm]{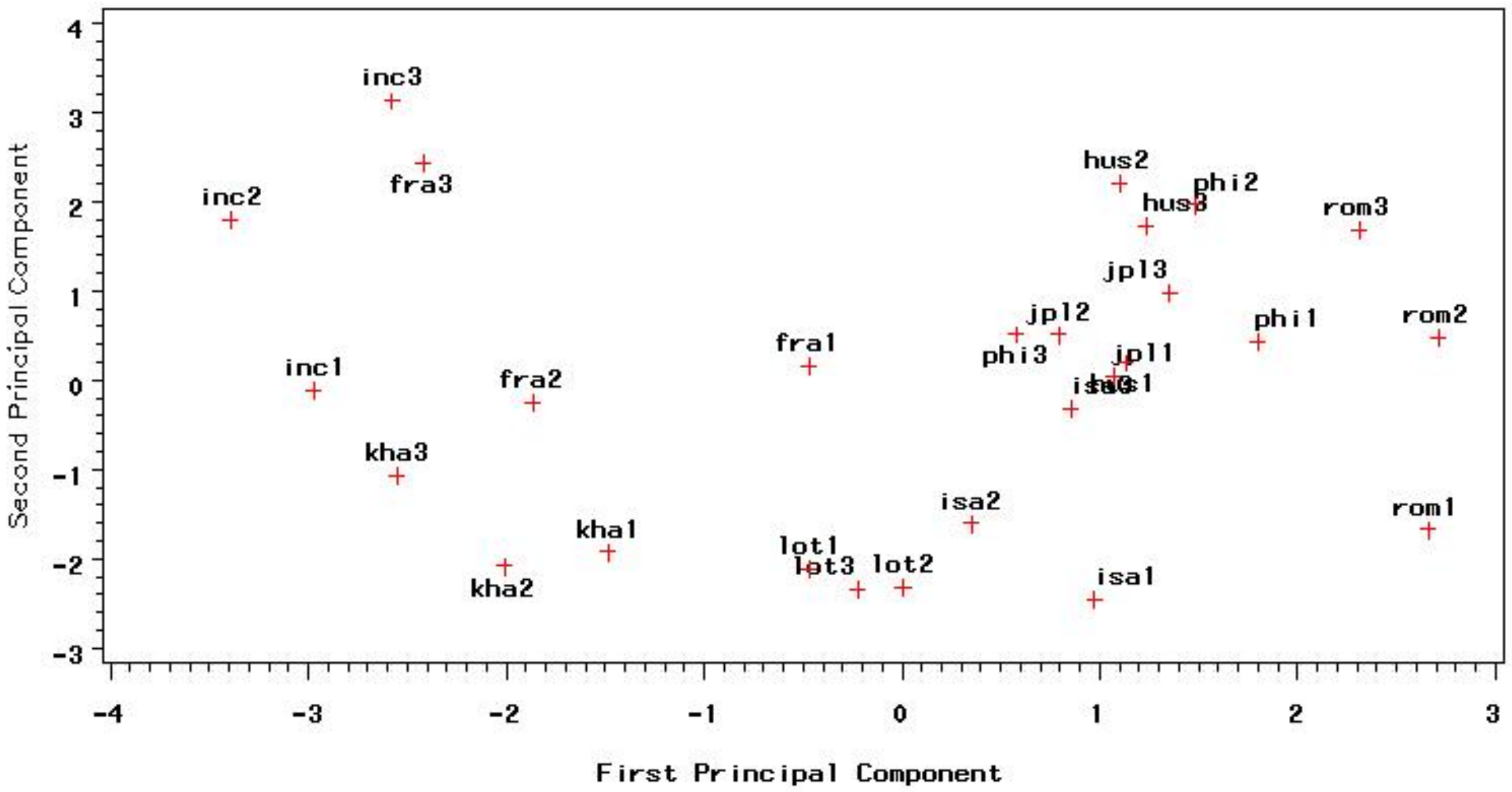} \\
(a) & (b)\\
\multicolumn{2}{c}{ \includegraphics[height=4cm,width=5.5cm]{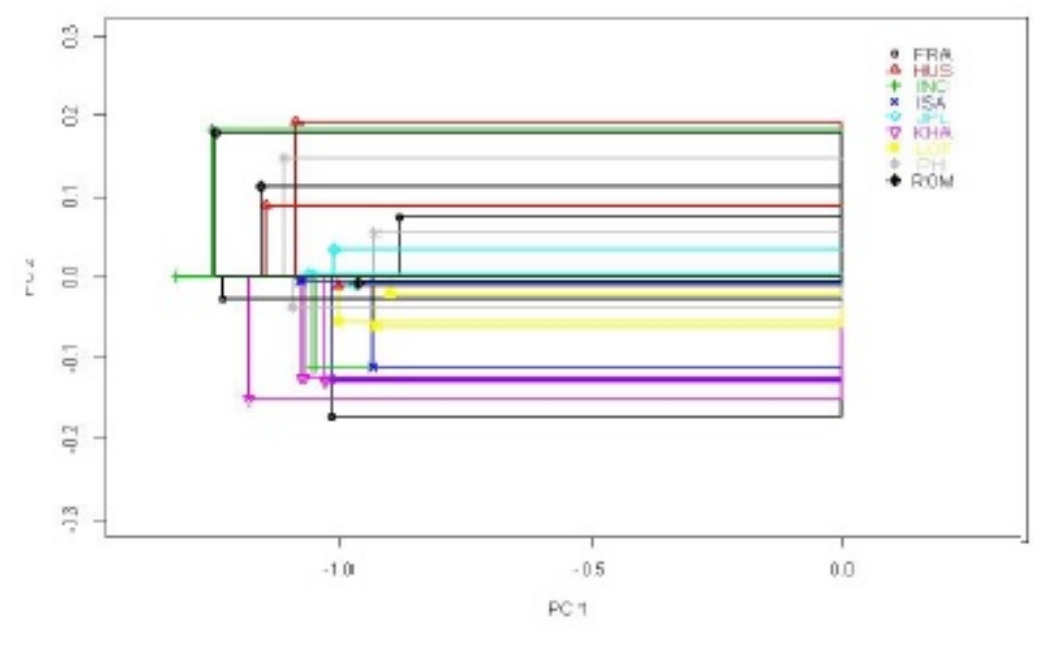}} \\
\multicolumn{2}{c}{ (c)}\\
\end{tabular}
\caption{Faces Data -  $PC_1~ \times PC_2$  - (a) Vertices (b) Ranges (c) Range transformation} \label{fig2}
\end{figure}

Figure $\ref{fig2}$ shows the corresponding principal component plots from these data when methods using the centers and /or ranges (or equivalently the end-points, i.e., vertices of the intervals) are used. Figure $\ref{fig2}$(a) (from Douzal et al., 2011, Fig. 6) shows the plots based on the vertices (interval end-points); Figure $\ref{fig2}$(b)  (from Douzal et al., 2011, Fig. 9) shows the plots when the ranges are used; and Figure $\ref{fig2}$(c) (from Le-Rademacher and Billard, 2012, Figure 5(a)) results from using the Lauro and Palumbo (2000) range-transformation method. These three methods all use classical surrogates in their varying ways. Comparing the plots in Figure $\ref{fig2}$ with that of Figure $\ref{fig1}$, we can see that these approaches cannot correctly classify the faces. This is particularly evident when comparing Figure $\ref{fig2}$(b) with Figure $\ref{fig1}$ in light of the additional knowledge that the faces are actually nine sets of three measurements from each of nine persons. We observe that the three faces (rom1, rom2, rom3, e.g.) in Figure $\ref{fig2}$(b) are not clustered, as they are in Figure $\ref{fig1}$; likewise, for some other faces, the classical approaches do not necessarily form the 3-wise clusters as would be expected.

Details of the associated analytic  diagnostics (such as inertia, etc.) including interpretations
 along with comparisons with PCA methodology based on classical surrogates can be found in Le-Rademacher and Billard (2012).

\section{Conclusion}

The initial theoretical work on deriving maximum likelihood (ML) estimators of parameters for interval data was that of  Le-Rademacher and Billard (2011).  Though    important,  their results were limited to obtaining MLE estimators for the mean and variance of a single interval-valued variable. However, the covariance statistic is a basic requirement for many methodologies (not  just for standard data but also for interval data), including in particular regression analysis, principal component analysis,  canonical correlation analysis,   among others.  Therefore, in this paper we have redressed  the  Le-Rademacher and Billard  limitation by extending their  results to deriving the MLEs for the core descriptive statistics for the two-dimensional case needed in these methodologies. The proposed MLE estimation approach  can be developed  for $p>2$-dimensional interval-valued variables   by  employing a $p$-dimensional normal distribution for the internal means, i.e.,  $\boldsymbol\Theta_{i1}$, and a $p$-variate Wishart distribution for the internal variations, i.e.,  $\boldsymbol\Theta_{i2}$,   in the proposed likelihood function.  The Le-Rademacher and Billard results emerge as special  cases of our wider derivations. Asymptotic properties of the proposed maximum likelihood estimators  are also derived.

\subsection*{Appendix A}

Early work on interval data sometimes transformed the interval-valued variable  into two variables, center and range (or, given their one-to-one correspondence equivalently into the end point values). Consider  the $Y$ values of the interval-valued data sets of Table $\ref{tab4}$.  Let us denote the interval centers by $Y^c = (a + b)/2$, and the interval half-range by $Y^r = (b - a)/2$. Then the first and   second columns of Table $\ref{tab5}$(a) give the sample variances of  $Y$ for the interval centers and ranges, respectively (calculated by classical results or as special cases of Eq.(2)). The third column shows the sum $(\mbox{Var}(Y^c) + \mbox{Var}(Y^r))$. This can be compared with the sample variance $\mbox{Var}(Y)$ of the intervals in the right-most column (from Eq.($\ref{eq2}$) and   Bertrand and Goupil, 2000). Thus we see that sometimes the sum $(\mbox{Var}(Y^c) + \mbox{Var}(Y^r))$ is greater, and sometimes less,  than the symbolic variance $\mbox{Var}(Y)$; this depends on the actual data. The fourth data set consists of classical values (with $a\equiv [a,a]$); in this case, the $\mbox{Var}(Y^r)=0$ and so  $\mbox{Var}(Y^c) = \mbox{Var}(Y)$, as it should.
\begin{table}[h] 
\caption{Some Data Sets ($Y, X$)} \label{tab4}
\centering
\scalebox{.86}{%
\begin{tabular}{|cc|cc|cc|cc|} \hline
\multicolumn{8}{|c|}{Data Sets }  \\  \hline
 \multicolumn{2}{|c|}{1}&\multicolumn{2}{|c|}{2}&\multicolumn{2}{|c|}{3}&\multicolumn{2}{|c|}{4} \\ \hline
$Y$ &$X$&$Y$ &$X$&$Y$ &$X$&$Y$ &$X$\\ \hline
[6,7]&[1,4]& [6,12]&[3,7]&[3,4]&[5,9]&[3, 3]&[4,4]\\
$[6,9]$&$[2,7]$&$[3,15]$&$[1,8]$&$[1,6]$&$[4,8]$&$[6,6]$&$[5,5]$\\
$[5,8]$&$[1,5]$&$[3,22]$ &$[2,15]$&$ [2,5]$&$[4,10]$&$[5,5]$&$[3,3]$\\
&&& &[2,4]&[3,7]& & \\ \hline
\end{tabular} }
\end{table}

 Likewise,   by using the  centers and range values for both $Y$ and $X$, we can calculate the classical covariances of the centers and of the ranges  and the symbolic interval covariances, from Eq.($\ref{eq3}$), shown in Table  $\ref{tab5}$(b). Again, the sum  $(\mbox{Cov}(Y^c, X^c) + \mbox{Cov}(Y^r, X^r))$ can be greater, or  smaller, than the  symbolic covariance $\mbox{Cov}(Y, X)$; and for classical observations, this sum equals the symbolic covariance correctly as expected.
 
\begin{table}[h] 
\caption{Variances and Covariances} \label{tab5}
\centering
\scalebox{.86}{%
\begin{tabular}{|c|ccc|c |} \hline
& \multicolumn{4}{c|}{ (a) - Variances $Y$}\\ \cline{2-5}
Set& $\mbox{Var}(Y^c)$&$\mbox{Var}(Y^r)$&$\mbox{Var}(Y^c)+\mbox{Var}(Y^r)$& $\mbox{Var}(Y)$\\ \hline
1& 0.222& 0.889&1.111 & 0.750\\
2& 2.722 & 28.222 & 30.944 & 17.750\\
3&0.047 & 2.188 & 2.234 & 0.859\\
4& 1.556 & 0.000 & 1.556 & 1.556\\ \hline \hline
& \multicolumn{4}{c|}{ (b) - Covariances ($Y,X)$}\\\cline{2-5}
Set& $\mbox{Cov}(Y^c, X^c)$&$\mbox{Cov}(Y^r, X^r)$&$\mbox{Cov}(Y^c,X^c)+\mbox{Cov}(Y^r,X^r)$& $\mbox{Cov}(Y,X)$\\ \hline
1& 0.389 & 0.667 & 1.056 & 1.222\\
2& 2.917 & 19.667 & 22.583 & 12.778\\
3& 0.156 & 0.125 & 0.281 & 1.198\\
4& 0.333 & 0.000 & 0.333 & 0.333\\ \hline
\end{tabular} }
\end{table}

For a second aspect, suppose a data set consists of intervals all with the same center but different range values. Then, the variance-covariance terms for the centers are zero; and in contrast, if the data are such that the observations have different center values but all have the same range value, then the variance-covariance terms for the ranges are zero. Then for  methods that rely on the relevant variance-covariance matrices,  the methodology cannot be properly implemented, since, e.g., in regression that matrix is zero and for principal components the eigenvalues are zero.

The variance-covariance definition of Eq.($\ref{eq3}$) does not have these limitations.

\subsection*{Appendix B}
The log likelihood function $\ln L_I$ from Eq.($\ref{eq13}$) and Eq.($\ref{eq14}$) is
\begin{align*}
\ln L_I &\propto -n \ln(\sigma_x)-n \ln(\sigma_y)-(n/2)\ln(1-\rho^2) \nonumber \\
    & ~~~ -\frac{1}{2(1-\rho^2)}\sum_{i=1}^n\bigg[ \frac{\left(\theta_{i1}^x -\mu_x\right)^2}{\sigma^2_x}
      + \frac{\left(\theta_{i1}^y -\mu_y\right)^2}{\sigma^2_y}
-2\rho\frac{(\theta_{i1}^x-\mu_x)(\theta_{i1}^y-\mu_y)}{\sigma_x\sigma_y}\bigg] \nonumber \\
&~~~ -\frac{n\nu}{2} \ln\left(\gamma_1\gamma_2-\gamma_3^2 \right)
-\frac{\gamma_1\gamma_2}{2(\gamma_1\gamma_2-\gamma_3^2)} \left( \frac{1}{\gamma_1}\sum_{i=1}^{n} \theta_{i2}^{x} +\frac{1}{\gamma_2}\sum_{i=1}^{n} \theta_{i2}^{ y} -\frac{2\gamma_3  }{\gamma_1 \gamma_2} \sum_{i=1}^{n} \theta_{i2}^{ xy}  \right)
\end{align*}

Then successively differentiating $\ln L_I$ with respect to each of the eight parameters in $\mbox{\boldmath$\tau$}$, we obtain
{\small
\begin{align}
  \frac{\partial \ln L_I}{\partial \mu_x} &= \frac{1}{(1-\rho^2) } \left( \frac{1}{\sigma^2_x} \sum_{i=1}^n\left( \theta_{i1}^x-\mu_x\right) -\frac{\rho}{\sigma_x \sigma_y}\sum_{i=1}^n\left(\theta_{i1}^y-\mu_y\right) \right),  \nonumber
  \\
  \frac{\partial  \ln L_I}{\partial \mu_y} &= \frac{1}{(1-\rho^2)} \left( \frac{1}{\sigma^2_y} \sum_{i=1}^n\left(\theta_{i1}^y-\mu_y\right) -\frac{  \rho}{\sigma_x \sigma_y}\sum_{i=1}^n\left(\theta_{i1}^x-\mu_x\right)\right),\nonumber 
  \\
  \frac{\partial  \ln L_I}{\partial \sigma_x} &= \frac{-n}{\sigma_x} + \frac{1}{2(1-\rho^2)}\left(\sum_{i=1}^n\frac{2\left(\theta_{i1}^x-\mu_x\right)^2}{\sigma^3_x}
  - 2\rho\sum_{i=1}^n\frac{\left(\theta_{i1}^x-\mu_x\right)\left(\theta_{i1}^y-\mu_y\right)}{\sigma_x^2\sigma_y}\right),  \nonumber 
  \\
  \frac{\partial  \ln L_I}{\partial \sigma_y} &= \frac{-n}{\sigma_y} + \frac{1}{2(1-\rho^2)}\left(\sum_{i=1}^n\frac{2\left(\theta_{i1}^y-\mu_y\right)^2}{\sigma^3_y}
  - 2\rho\sum_{i=1}^n\frac{\left(\theta_{i1}^x-\mu_x\right)\left(\theta_{i1}^y-\mu_y\right)}{\sigma_x\sigma_y^2}\right),  \nonumber  
  \\
  \frac{\partial  \ln L_I}{\partial \rho} &= \frac{n\rho}{(1-\rho^2)} - \frac{\rho}{(1-\rho^2)^2}
  \left(\sum_{i=1}^n \frac{ \left( \theta_{i1}^x-\mu_x\right)^2}{\sigma^2_x}  + \sum_{i=1}^n \frac{ \left(\theta_{i1}^y-\mu_y \right)^2}{\sigma^2_y}  \right) \nonumber \\
   &~~~~~~~~~~~~~~~~~~~~~~~~~~~~~~~~~~~~~ + \frac{1+ \rho^2}{(1-\rho^2)^2} \sum_{i=1}^n\frac{(\theta_{i1}^x-\mu_x)(\theta_{i1}^y-\mu_y)}{\sigma_x\sigma_y},    \label{eq20}
    \\
  \frac{\partial  \ln L_I}{\partial \gamma_1} &= -\frac{n\nu \gamma_2}{2G} + \frac{\gamma_2^2}{2G^2}\sum_{i=1}^n\theta_{i2}^x  - \frac{G-\gamma_1 \gamma_2}{2G^2}\sum_{i=1}^n\theta_{i2}^y - \frac{\gamma_2\gamma_3}{G^2} \sum_{i=1}^n\theta_{i2}^{xy},\nonumber  
  \\
 \frac{\partial  \ln L_I}{\partial \gamma_2} &= -\frac{n\nu \gamma_1}{2G} - \frac{G-\gamma_1 \gamma_2}{2G^2}\sum_{i=1}^n\theta_{i2}^x + \frac{\gamma_1^2}{2G^2}\sum_{i=1}^n\theta_{i2}^y    - \frac{\gamma_1\gamma_3}{G^2} \sum_{i=1}^n\theta_{i2}^{xy},  \nonumber 
  \\
  \frac{\partial  \ln L_I}{\partial \gamma_3}& =  \frac{n\nu \gamma_3}{ G} - \frac{ \gamma_2 \gamma_3}{G^2}\sum_{i=1}^n\theta_{i2}^x   - \frac{ \gamma_1 \gamma_3}{G^2}\sum_{i=1}^n\theta_{i2}^y   + \frac{ G+ 2\gamma_3^2}{ G^2}\sum_{i=1}^n \theta_{i2}^{xy} \nonumber 
\end{align}}
where $G=\gamma_1\gamma_2 - \gamma_3^2$.

Then, substituting the relevant maximum likelihood estimator and setting the derivatives to zero, we can obtain the maximum likelihood estimators $ \hat{\mbox{\boldmath$\tau$}}_{xy} = (\hat{\mu}_x, \hat{\mu}_y, \hat{\sigma}_x^2, \hat{\sigma}_y^2, \hat{\rho}, \hat{\gamma}_1,\hat{\gamma}_2,\hat{\gamma}_3)$ for $\mbox{\boldmath$\tau$}_{xy} = (\mu_x, \mu_y, \sigma_x^2, \sigma_y^2, \rho, \gamma_1,\gamma_2,\gamma_3)$ to be as given by Eq.($\ref{eq29}$)-Eq.($\ref{eq32}$).  We also note that instead of solving the partial derivative in Eq.($\ref{eq20}$) for the derivation of the estimator $\hat{\rho}$, we can more easily obtain the result of Eq.($\ref{eq20}$) by following, e.g., Casella and Berger (2002, p.358) who suggest using a partially maximized likelihood function.



\end{document}